\documentclass[twoside]{ilcws07}
\usepackage[latin1]{inputenc}
\usepackage[dvips]{graphicx,epsfig,color}
\usepackage{wrapfig,rotating}
\usepackage{amssymb,amsmath,array}
\usepackage{wrapfig}

\begin{document}
\title{The Detector DCR} 
\author{Akiya Miyamoto
\thanks{Representing co-editors: Ties Behnke, Chris Damerell and John Jaros}
\vspace{.3cm}\\
High Energy Accelerator Research Organization (KEK),\\
1-1 Oho, Tsukuba, Ibaraki 305-0801, Japan
}

\maketitle

\begin{abstract}
The Detector Concept Report(DCR) consists of two parts, one for the physics and the 
other for ILC detectors. It has been prepared as the accompany document 
of the ILC Reference Design Report. The overview of the 
detector part of the DCR and the plan for the final release is presented in this talk.
\end{abstract}

\section{Introduction}
The preparation of the DCR has been started since LCWS2006 at Bangalore\cite{LCWS2006}.
Four editors for the detector part, Ties Behnke, Chris Damerell, John Jaros and Akiya Miyamoto,
have worked together with authors of sub-sections to prepare the document.  
The preliminary version has been open to the community after the 
workshop at Beijing (BILCW07)\cite{BILCW07}.
Taking into account comments from the community as well as those from the Review Panel,
it is scheduled to be released in August this year\cite{DetectorRDR}.

The goal of the Detector DCR is to make the case that detectors can do the ILC physics, 
showing detector designs are within our reach,
where we are in detector developments and where we are going.
On the other hand, the DCR is neither a complete description of a detector 
nor a review of the ILC detector concepts.
The detector DCR is based on Detector Outline Documents (DODs)\cite{gld, ldc, sid, fourth}  prepared by 
four detector concept teams last year as well as  new studies
since then, but a little focus is put on concept specific issues.

Selected topics of the detector DCR is presented in the next section and 
the plan for the final release is described in the subsequent section.

\section{Overview of the Detector DCR}

The goal of the ILC physics includes understanding of the mechanism of mass generation and electroweak 
symmetry breaking, searching for and perhaps discovering supersymmetric 
particles and confirming the principle of supersymmetry, and 
hunting for signs of extra space-time dimensions and quantum gravity\cite{zervas}.
The ILC detectors have to be optimized for these ILC physics targets.

Experimental conditions at the ILC provide an ideal environment for the precision study of 
elementary particle interactions, thanks to the clean signal conditions and well-defined
initial state.  Events are recorded without a bias which might be caused by an event trigger.
However, the physics poses challenges on detector performances,
pushing the limits of jet energy resolution, tracker momentum resolution,
and vertex impact parameter resolution, as well as full solid angle coverage.
Although benign by LHC standards, the ILC environment poses some 
interesting challenges of its own.

The world-wide linear collider physics and detector community has worked on these 
challenges and made impressive progresses.
Four teams, GLD\cite{gld}, LDC\cite{ldc}, SiD\cite{sid} and 4th\cite{fourth},
have formed to study detector concepts for the ILC experiments.
They have reported their studies as the Detector Outline Documents (DODs)
last year, and have kept continuing concept studies.
GLD, LDC, and SiD are equipped with a granular calorimeter 
for particle flow measurements, while 4th aims to achieve 
a good jet energy resolution by a dual-readout calorimeter.
Key parameters of the four detector concepts are summarized in Table~\ref{Detectors:concepts}.

\begin{table}[hbt]
\begin{center}
\begin{minipage}{0.90\textwidth}
\caption{{\it Some key parameters of the four detector concepts.  See Table~\ref{tab-magnet} 
for magnet parameters.}
\label{Detectors:concepts}}
\end{minipage}

\begin{small}

\begin{tabular}{| l | c | c | c | c | } \hline
 & GLD & LDC & SiD & 4th \\ \hline
VTX & pixel & pixel & pixel & pixel \\
~~ $R_{in}/R_{out}$ (cm) & 2.0/5.0 & 1.6/6.0 & 1.4/6.1  & 1.5/6.1 \\ \hline
Main Tracker & TPC$\left[{\rm Si}\right]$ & TPC$\left[{\rm Si}\right]$ & Si & TPC(drift) \\
~~ $R_{in}/R_{out}$(TPC$\left[{\rm Si}\right]$) (cm) & 45/200$\left[9/30\right]$ & 30/158$\left[16/27\right]$ & 20/127 & 20/140 \\
~~ $L_{half}$(TPC$\left[{\rm Si}\right]$) (cm) & 230$\left[62\right]$ & 208$\left[140\right]$ & 168 & 150 \\
~~ \# barrel points (TPC$\left[{\rm Si}\right]$) & 200$\left[4\right]$ & 200$\left[2\right]$ & 5 & 200(120) \\ \hline
ECAL & Scinti.-W & Si-W & Si-W & Crystal \\
~~ Barrel $R_{in}/L_{half}$ (cm) & 210/280 & 160/230 & 127/180 & 150/240 \\
~~ \# $X_0$ & 27 & 23 & 29 & 27 \\ \hline
HCAL & Scinti.-Fe & Scinti.-Fe & RPC/GEM-W & fiber Dream \\
~~ Barrel $R_{in}/L_{half}$ (cm) & 229.8/280 & 180/230 & 141/277.2 & 180/280 \\ 
~~ Interaction length & 5.8 & 4.6 & 4.0 & 9 \\ \hline
Overall Detector & & & & \\
~~~ $R_{out}/L_{half}$ (cm) & 720/750 & 600/620 & 645/589 & 550/650 \\ \hline
\end{tabular}
\end{small}

\end{center}
\end{table}

In parallel to the concept studies, 
R\&D on detector technologies have been pursued actively 
world-wide\cite{detector-RandD-panel}. Inter-concept teams 
have been formed to address R\&D issues common to concepts.

The detector DCR is based on these activities, but with a little emphasis on 
concept specific issues.

\subsection{Challenges for Detector Design and Technologies}
\label{detector-challenge}

The relatively low radiation environment of the ILC allows detector designs and
technologies not possible at the LHC, but the demanding physics goals 
still challenge the state of the art technologies.

\begin{wrapfigure}{r}{0.66\textwidth}
\centerline{\includegraphics[width=0.60\textwidth]{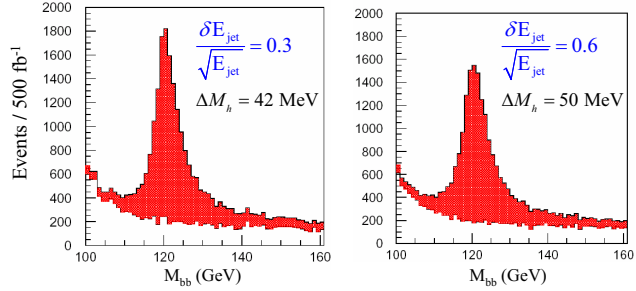}}
\begin{small}
\caption{{\it Reconstructed Higgs di-jet invariant mass for different 
jet energy resolutions. The analysis has been performed for a center of mass energy 
of 350 GeV and a total integrated luminosity of 500 fb $^{-1}$}}\label{fig-higgs-4jet-mass}
\end{small}
\end{wrapfigure}

Many of the interesting physics processes at the ILC appear in multi-jet final states, often 
accompanied by charged leptons or missing energy. The reconstruction of the invariant mass of 
two or more jets will provide an essential tool for identifying and distinguishing $W$'s, $Z$'s, 
H's, top and discovering new particles.To distinguish $W$'s and $Z$'s in their hadronic decay 
mode, the di-jet mass resolution should be comparable to their natural width, say a few GeV or 
less. The jet energy resolution of $\sigma_E/E < 3 \sim 4\%$ ( $30\%/\sqrt{E}$ for jet energies 
below about 100 GeV),
which is about a factor of two better than that achieved at LEP,
will provide such di-jet mass resolution. 
A factor of two improvement in jet energy measurement 
improves the resolution of the Higgs mass measurement using
the four-jet mode of the Higgsstrahlung process by about 20\% 
as shown in Figure~\ref{fig-higgs-4jet-mass}.
It is equivalent to a luminosity gain of about 40\%.
A similar gain of performance is expected in measurements of 
such as $\Delta$Br($H\rightarrow WW^*$)
and the Higgs self-coupling.

The Higgs measurement from di-lepton recoil mass is important
because it is measured without any assumption on its decay mode.
In order to measure the Higgs mass at a precision close to 
the ultimate limit set by the initial beam energy spread, 
the momentum resolution of the tracking system ($\Delta p_t/p_t$) has to be 
less than $1\times 10^{-3} \oplus 5\times 10^{-5} p_t$(GeV/c).
Such a high-performance tracking device allows measurements of the center of mass energy 
at about 20 MeV precision by using the process,
$e^+e^- \rightarrow \mu^+\mu^-(\gamma)$.  
In the measurement of the slepton mass using the end point of 
lepton momentum, a gain of about 40\% in luminosity is expected 
if the momentum resolution improves from 
from $8\times 10^{-5}p_t$ to $2\times 10^{-5} p_t$.

Efficient and clean identification of bottom and charm quark jets 
are indispensable methods to carry 
out the ILC physics program.  For example, the identification of 
$b$ and $c$ jets in Higgs decays are essential to measure Yukawa 
couplings of $c$ and $b$ quarks.  $b$ jets identification in 
the top quark decays are useful to reduce combinatorial 
background in finding a correct jet combination of 
their hadronic decay.  
Quark charge measurements of jets through an efficient 
reconstruction of secondary and thirdly vertices would be 
a key method for studies of forward-backword assymetries of $b$ quark.
The vertex detector which could measure 
the impact parameter at precision better than 
$5 \oplus 10 / p \sin^{3/2}\theta$ ($\mu m$)
will provide the performance to carry out these physics.

Sub-detector performances needed for key ILC physics measurements are summarized 
in Table~\ref{tab-performance-requirement}.
\begin{table}[hbt]
\begin{center}
\caption{Sub-Detector Performance Needed for Key ILC Physics Measurements.
\label{tab-performance-requirement}}
\begin{scriptsize}
\begin{tabular}{|l@{\hspace{-4pt}}|l@{\hspace{-4pt}}|c@{\hspace{-4pt}}|c@{\hspace{-4pt}}|c|}
\hline
Physics Process & Measured Quantity 
 & \begin{tabular}{@{\hspace{-6pt}}c} Critical \\ System  \end{tabular}
 & \begin{tabular}{@{\hspace{-6pt}}c} Critical Detector \\ Characteristic \end{tabular}
 & \begin{tabular}{@{\hspace{-6pt}}c} Required \\ Performance  \end{tabular} \\ \hline

 \begin{tabular}{@{\hspace{0pt}}l} $ZHH$ \\ $HZ \rightarrow q\bar{q}b\bar{b}$ 
  \\ $ZH \rightarrow ZWW^*$  \\ $\nu\bar{\nu} W^+W^-$ \end{tabular}
& \begin{tabular}{@{\hspace{0pt}}l} Triple Higgs Coupling \\ Higgs Mass 
  \\  B($H\rightarrow WW^*$) \\  $\sigma({e^+e^- \rightarrow \nu\bar{\nu} W^+W^-})$ \end{tabular} 
& \begin{tabular}{@{\hspace{0pt}}c} Tracker \\ and \\ Calorimeter \end{tabular} 
& \begin{tabular}{@{\hspace{0pt}}c} Jet Energy \\ Resolution, \\ $\Delta E/E$ \end{tabular} 
& $3 \sim 4$ \% 
\\ \hline

\begin{tabular}{@{\hspace{0pt}}l} $ZH \rightarrow \ell^+\ell^-X$ \\ $\mu^+ \mu^- (\gamma)$ 
  \\ $HX \rightarrow \mu^+\mu^- X$ \end{tabular} 
& \begin{tabular}{@{\hspace{0pt}}l} Higgs Recoil Mass \\ Luminosity Weighted $\rm E_{cm} $ 
  \\ B($H\rightarrow \mu^+ \mu^-$) \end{tabular} 
& Tracker\hspace*{24pt}
& \begin{tabular}{@{\hspace{0pt}}c} Charged Particle \\ Momentum Resolution,
  \\ $\Delta p_t/p_t^2$  \end{tabular} 
& $5\times 10^{-5}$ 
\\ \hline

\begin{tabular}{@{\hspace{0pt}}l} $HZ, H \rightarrow b\bar{b},c\bar{c},gg$ 
  \\ $b\bar{b}$ \end{tabular} 
& \begin{tabular}{@{\hspace{0pt}}l} Higgs Branching Fractions \\ $b$ quark charge asymmetry \end{tabular} 
& \begin{tabular}{@{\hspace{0pt}}c} Vertex \\ Detector \end{tabular} 
& \begin{tabular}{@{\hspace{0pt}}c} Impact \\ Parameter, $\delta_b$ \end{tabular} 
& \begin{tabular}{@{\hspace{0pt}}c} $ 5 \mu \mathrm{m} ~\oplus $ 
    $10 \mu \mathrm{m}$ \\ $/ p({\mathrm GeV/c})\sin^{3/2}\theta$ \end{tabular} 
\\ \hline

\end{tabular}
\end{scriptsize}
\end{center}
\end{table}

\subsection{Machine Detector Interface}


The ILC beam induces following backgrounds; disrupted beam,
photons and low energy electron-positron pairs generated by beamstrahlung;
synchrotron radiation created when beam pass through 
beam line magnets; muons created by interactions between beam halo and collimators;
neutrons created by electron-positron pairs and disrupted beam hitting beam line components;
hadrons and muons created by photon-photon interactions.

A careful design of shields against these backgrounds is 
crucial. Their impacts on detector performances have been 
studies based on Monte Carlo simulations and  estimated background hit rates 
have been below critical level so far. For an example, 
a hit occupancy in TPC due to the electron-positron pair background 
has been estimated by a simulation.
TPC takes 100 bunch crossing(BX) of time to readout an event.
After superimposing 100 BX of background hits, 
the hit occupancy is less than 0.2\%, which is well 
below the critical occupancy of 1\%.

Concerning the detector integration, the baseline plan is 
to assemble most of the detectors on surface, then brought them 
down the underground experimental fall for final assembly.
This is to minimize the size of the underground experimental hall 
and to save the detector construction time.

The baseline design of the ILC foresees one interaction region, equipped with two detectors.
The two detectors are laid out in such a way that each can be moved quickly 
in and out the interaction region thus allowing the sharing of luminosity between 
both detectors (push-pull operation).
Details such as switchover time and frequency are still under discussion 
and a system with two beam delivery lines will be kept as an 
option until the detailed engineering design study demonstrates 
the feasibility of such a push-pull scheme.

\subsection{Subsystem Design and Technologies}
Technologically oriented description of detector sub-systems for a ILC detector is 
described in this section, aiming to show 
what kind of technologies exists for them, their challenges,
and required R\&Ds to achieve goals.

\subsubsection{Vertex Detector}
Four to six layers of silicon pixel detectors are used for a vertex detector.
In total there are about $10^9\sim 10^{10}$ pixels of size of about 20 $\mu$m$^2$ or less.  The beam pipe radius 
is 15 mm or less to place the vertex detector as close as to the interaction point. 
The thickness of each layer of the vertex detector is 0.1\% $X_0$ or less.  The vertex detector has to be 
reasonably hard against radiation and beam induced RF radiation (EMI).
To keep background hits occupancy low, it has to be readout our fast or store locally 
and readout between the beam pulse.
Due to a unique feature of the ILC beam structure, which has about 200 msec of quiet period after 1 msec 
of beam collisions, data of all collisions have to be read out without
a front-end trigger for software filltering at later stages.
To reach the performance goal, a calibration of internal alignment has to be carefully designed
and an effect of powering and cooling to detector allignments should be minimum.
There are no proven vertex detector technology to meet the performance 
goal under the ILC operational condition and R\&Ds on more than 10 technologies are pursued 
worldwide extensively.

\subsubsection{Silicon Strip Tracker}
Silicon strip tracker is used as the main tracker of SiD concept and intermediate, forward
or endcap trackers of other concepts.
The silicon strip tracker is robust against unexpected radiation backgrounds;
it is fast such that signal charges are collected before the next bunch crossing and 
an impact of beam backgrounds are minimum;
it is precise such that point resolutions of $5\sim 10 \mu$m are achievable.
While silicon strip detector has been used extensively in other experiments,
large detector system has typically 2\%$X_0$ of material per layer.
The most of them is attributable to dead material needed for 
support, cooling and readout.  This dead material is 
a source of a peformance deterioration. To significanly 
reduce these dead material while keeping the benefits of slicon strip 
detectors is one of the most significant 
challenges of R\&Ds for silicon tracking at the ILC\cite{DetectorRDR}.

\subsubsection{Gaseous Tracker}
Time Projection Chamber (TPC) is considered as a main tracker by GLD, LDC and 4th concepts.
The tracking of the TPC is robust because of many three-dimensional point measurements along the track.
Material in the tracking volume is minimum and particle identification is possible.
Detectors such as GEM\cite{gem}
and MicroMegas\cite{micromegas} are candidates for the endplate 
detector, in order to meet the goal of the momentum resolution, 
Beam tests of a small test system
suggest that the performance goal is within the technology in hand. 
Still a design to minimize a positive ion build up in the drift volume has to be developed and
a gas with lowest diffusion and less contamination of Hydrogen atoms should be investigated.
Operatability in non-uniform magnetic field caused by the anti-DID magnet
and design of end-plate electronics with short radiation length is another challenge of the TPC R\&D.
International collaboration, LCTPC\cite{LCTPC}, is formed and pursing these studies.

\subsubsection{Calorimeter}
Calorimeter is a key device to achieve a good jet energy resolution.
The GLD, LDC and SiD concepts are equipped with  a particle flow calorimeter,
which is characterized by a highly granular segmentation both in lateral 
and longitudinal directions. A sandwitch structure of absobers and 
small sensors are adopted.  Both electromagnet and hadron calorimeters 
are placed in side the coil of the detector solenoid magnet.
In the particle flow analysis, charged particle signals 
in the calorimeter are set aside by using tracker information,
and calorimeter information is used only to measure neutral particle 
energies. Therefore, the high granularity in calorimeter segmentation and an excellent 
shower reconstruction alogirthm are crucial.
On the other hand, 4th concept is equipped with a dual read out calorimeter:
scintillating fibers for all charged particles in a shower
and clear fibers for Cherenkov light induced by electrons and positrons.
Despite it's few longitudinal sampling, 
it aims at a good jet energy resolution with a high resolution calorimeter.

A development of calorimeter technologies is one of the most active area
of the ILC detector R\&D\cite{DetectorRDR} and many technologies are currently 
pursed, for example; for electromagnetic calorimeter,  sandwiches of 
tungsten or lead absorber and silicon, MAPS, or scintillator and
semiconductor photon sensor readout; for hadron calorimeter, 
lead or iron as absorber and scintillator and photon sensor readout, 
gas chamber and GEM or RPC readout.  

\subsubsection{Superconducting Magnet}
A detector magnet is one of the major part of the detector cost.
The GLD, LDC and SiD concepts use a large bore coil, while 4th concept
use a dual coil system where the outer coil is used instead of iron flux return.
Typical parameters of them are summarized 
in Table~\ref{tab-magnet}.  As seen in the table, the parameters 
of the magnet for the ILC detector is similar to the CMS magnet and it's experience 
is useful.
\begin{table}
\begin{small}
\caption{{\it Summary of the parameters of ILC detector magnet, compared with that of CMS.}}\label{tab-magnet}
\begin{center}
\begin{tabular}{| l |  l || l || l | l | l | l | } \hline
 & unit & CMS  & GLD & LDC & SiD & 4th(In/Out)  \\ \hline
Magnetic Field & Tesla & 4 & 3 & 4 & 5 & 3.5/1.5 \\
Coil Radius & m & 3.25 & 4 & 3.16 & 2.65 & 3/4.5 \\
Coil Half length & m & 6.25 & 4.43 & 3.3 & 2.5 & 4/5.5 \\
Stored Energy(E) & GJ & 2.6 & 1.6 & 1.7 & 1.4 & 5.7 \\
Cold mass (M) & ton & 220 & 78 & 130 &  117 & \\ 
E/M & kJ/kg & 12.3 & 20 & 13 & 12 & \\ \hline
\end{tabular}
\end{center}
\end{small}
\end{table}

\subsubsection{Data Acquisition}
The ILC RF system is operated at the frequency of 5Hz.  During the beam period of 1 msec, 
the collision rate is about 3 MHz. A pipeline system is mandatory 
to record data of all collisions. The burst collision is followed by 
about 200 msec of a quiet time.  Thus average event rate is about 15kHz, 
which is moderate compare to LHC. No hardware trigger is planned
and event selection is done by software after readout data of all bunch collisions.  On the other hand,
zero suppression and data compression at detector front ends
are importantl to minimize a load to the data acquisition system, because
the ILC detectors are equipped with high granularity sensors.

\subsubsection{Luminosity, Energy, and Polarization}
The beam energy should be know to be less than 100 ppm precission 
for the precise Higgs recoil mass measurement.
For physics at GigaZ or $W$ threshold, it is required to be less than 50 ppm.
About 200 ppm has been achieved at LEP and SLC.  Several R\&Ds\cite{DetectorRDR} are in progress 
to achieve a factor of 2 or more improvement. These R\&Ds include 
the studies on developments of a high precision beam position monitor 
to measure beam energy using upstream beam line magnets as a spectrometer; 
the beam energy measurement by detecting synchrotron lights emitted 
from downstream beam line bending magnets; and the measurement of 
the energy weighted luminosity from lepton's acollinearity 
of processes such as Bhabha and $\mu\bar{\mu}(\gamma)$.

Beam polarization should be measured at precision better than 0.5\%.
A gain in physics potential is anticipated if $\Delta P \sim 0.25\%$ or less.
It is measured by Compton polarimeters at upstream and down stream of IP.
Developments of the instruments for the Compton measurements is important.

\subsubsection{Test beams}
Detector R\&D requires supports by  test beam resources.
Resources are limited and optimal coordination world wide is necessary.
Test Beams working group has been organized by WWS and the first report has been 
presented\cite{TestBeamReport}.

\subsection{Sub detector performance}
Each concept team has developed 
their own detector full simulator and reconstruction tools and pursued studies on 
performances of such as vertexing, tracking, jet reconstruction and so on.
It is impossible to cover all results here and only typical ones are shown.
Performances are more or less similar among the concepts.

The tracking performance has been studied for both  TPC and Silicon 
main tracker.  For the TPC main tracker, the track finding efficiency has been studied 
using $Z$ pole events where $Z$ decays to $d\bar{d}$. The obtained 
the efficiency exceeded 99\%, though 
realistic effects such as those by a non-uniform magnetic field, space charges and background hits 
have yet to be taken into account. 
SiD adopts an inside-out tracking finding method,
where the vertex detector is used to find a seed track.  According to this method, 
the efficiency of about 99\% is achived for a track whose origin is 
within 1 cm from the IP using a sample of $e^+e^- \rightarrow Z \rightarrow q\bar{q}$ events 
at 500 GeV center-of-mass energy. 
The momentum resolution of the tracking device has been studied by the GLD.
Combining information of TPC, the intermediate tracker and the vertex detector, 
the momentum resolution is found to be consistent with the gaol of 
$\Delta p_t / p_t \sim 10^{-3} \oplus 5 \times 10^{-5} p_t$ (GeV/c).

Impact parameter resolutions of the tracking system have also studied by 
each concept teams and found to be consistent with the performance goal.

\begin{wrapfigure}{r}{0.55\textwidth}
\centerline{\includegraphics[width=0.50\textwidth]{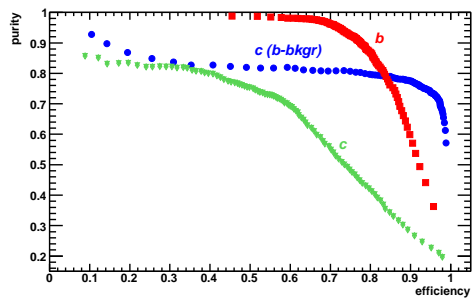}}
\begin{small}
\caption{{\it Efficiency and purity for 
tagging a $b$-quark(red square) and
$c$-quark(green triangle) jets in Z decays, using a full simulation.
The blue-circle points indicate the further improvement in 
performance of the charm tagging in events with only bottom 
background is relevant.}}
\label{fig-sec6-vtx-tageff}
\end{small}
\end{wrapfigure}

As already pointed out in the subsection~\ref{detector-challenge}, the pure and efficient tagging of 
$b$ quark and $c$ quark jets is important for the ILC physics.
The topological vertexing as pioneered by SLD has the
potential for sucha high performance tagging. The code has been ported for studies of ILC detectors.
An initial result of its study is shown in 
Figure~\ref{fig-sec6-vtx-tageff}\cite{ref-sonia-LCFIVertex}.
The obtained purity and efficiency using a realistic detector model is promising. 

GLD, LDC and SiD all utilize sampling calorimeters, whose energy resolution is essentially 
determined by the sampling fraction.  For single particles, the energy resolution 
of the electromagnetic calorimeters ranges from 14 to 17\%$/\sqrt{E}$ 
for the stochastic term and those for the hadron calorimeter ranges 
from 50 to 60\%$/\sqrt{E}$.  
For jet energy measurements, the particle flow analysis (PFA) is crucial 
to achieve the required level of performance. At the ILC detectors, 
the trackers can measure charged particles better than the calorimeters.
Thus, in the PFA, tracker signals are used to get charged 
particle information and calorimeter signals are used only 
to reconstruct neutral particles.  Since calorimeter is sensitive 
to charged particles as well, it is essential to develop
a sophisticated algorithm to fully utilize the fine granularity of the
calorimeters, identify and remove the calorimeter signals produced by charged particles.

\begin{figure}
\begin{minipage}{0.48\textwidth}
\centerline{\includegraphics[width=0.7\columnwidth]{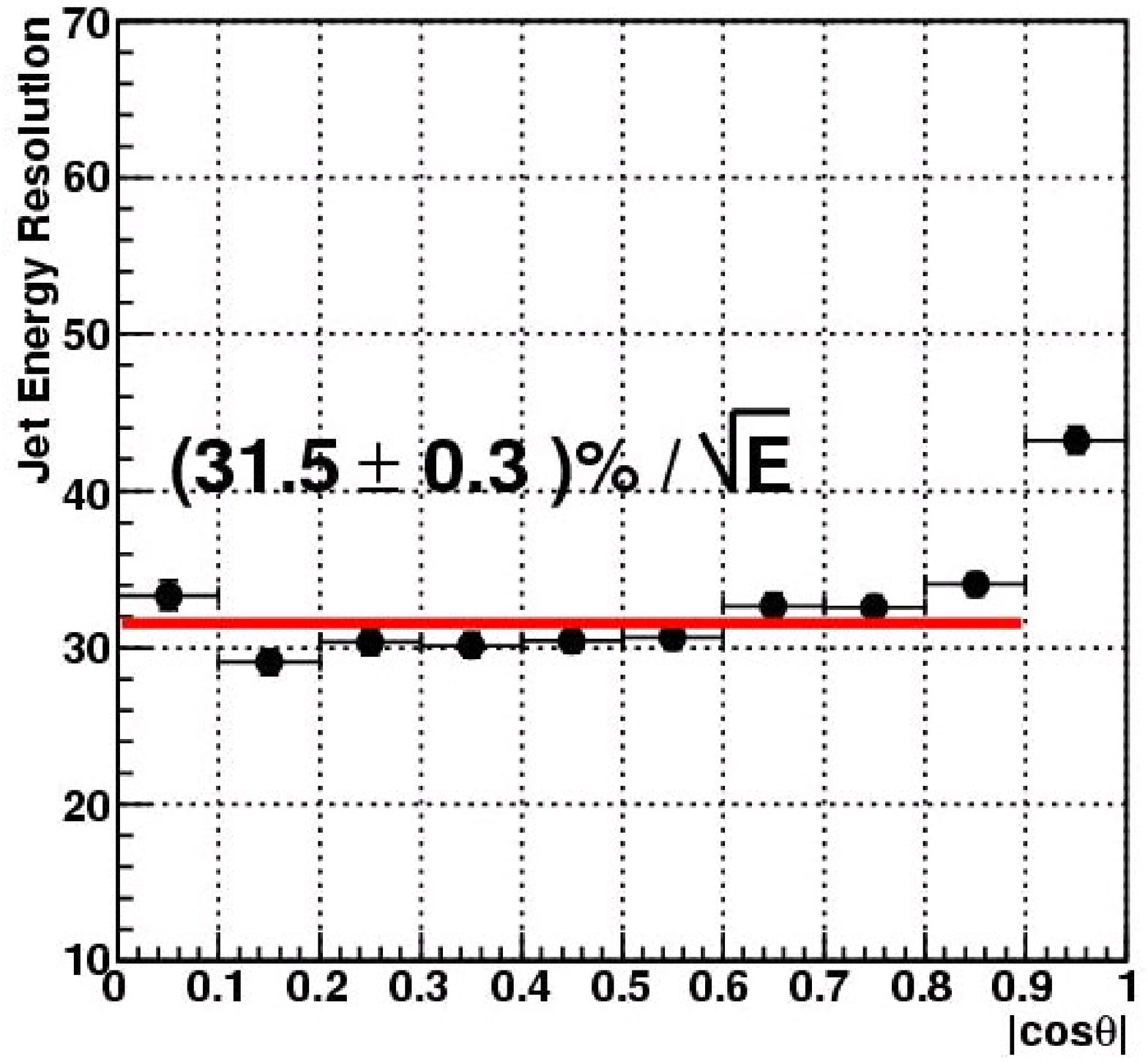}}
\begin{small}
\caption{{\it The stochastic term of the jet energy resolution ( $\sigma_{90}/\sqrt{E_{jet}}$)
as a function of $|\cos\theta_{jet}|$ in the case of $e^+e^-\rightarrow q\bar{q}$ (light quarks only)
events at $Z$ pole energy. A result by GLD-PFA for the GLD detector.}}
\label{fig-gldpfa}
\end{small}
\end{minipage}
\hspace{0.01\textwidth}
\begin{minipage}{0.48\textwidth}
\centerline{\includegraphics[width=0.7\columnwidth]{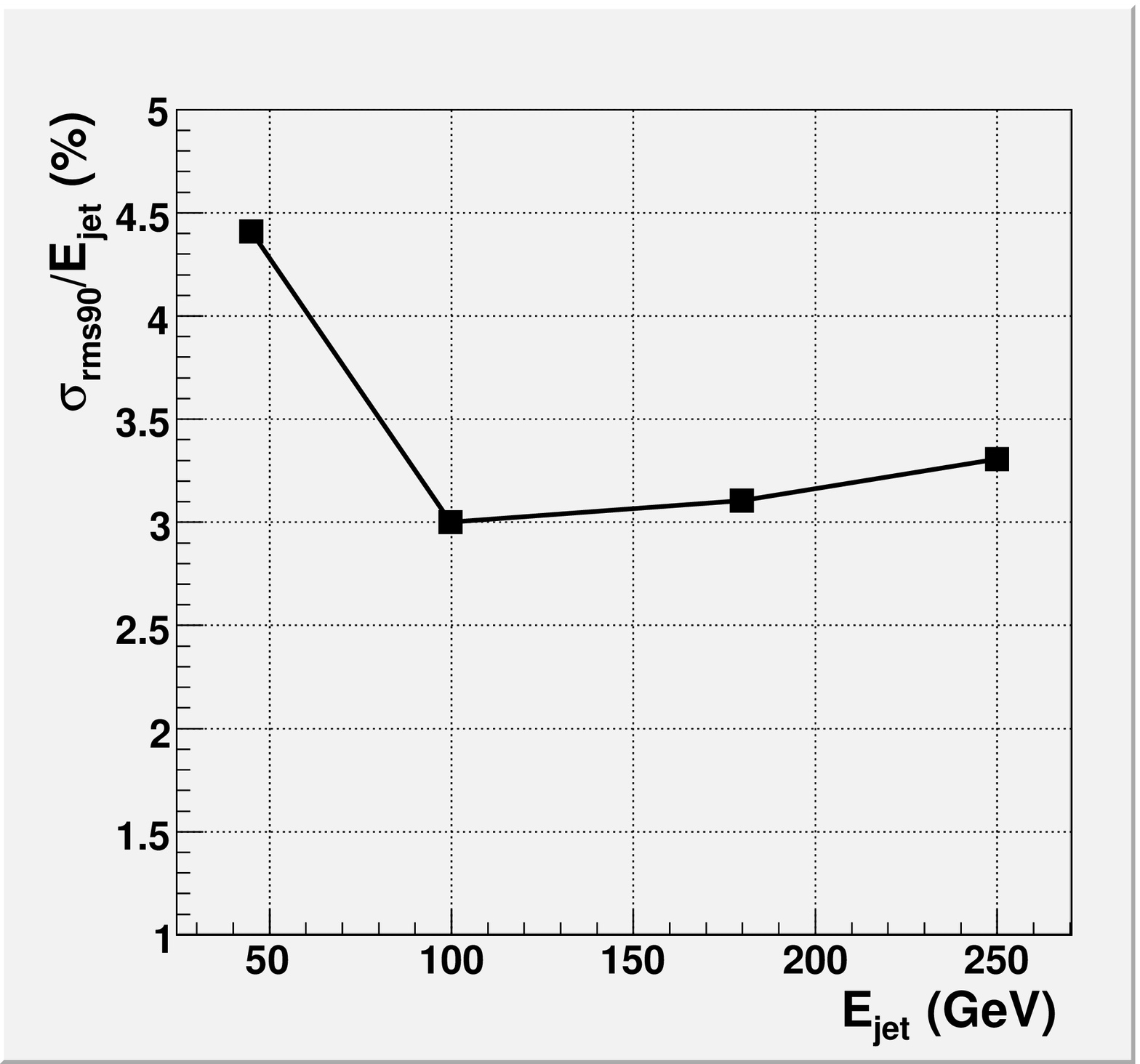}}
\begin{small}
\caption{{\it The relative jet energy resolution, $\sigma_{90}/E_{jet}$, 
of PandoraPFA averaged in the region $|\cos\theta_{jet}|<0.7$, as a function of the jet energy.}}
\label{fig-pfacompare}
\end{small}
\end{minipage}
\end{figure}

\begin{wrapfigure}{r}{0.50\textwidth}
\centerline{\includegraphics[width=0.48\textwidth]{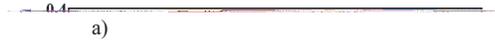}}
\begin{small}
\caption{{\it Jet energy resolution in terms of $\sigma_{90}/\sqrt{E}$ 
obtained with PandoraPFA and the Tesla TDR detector model 
plotted as a function of TPC outer radius and magnetic field.}}
\label{jetnergy-vs-tpcradius}
\end{small}
\end{wrapfigure}
To this end, PFA algorithm have been studied extensively by many
groups.  For an example, the algorithm such as WolfPFA\cite{wolfpfa} and GLD-PFA\cite{gldpfa} 
consists of following steps; cluster calorimeter signal cells; 
discard clusters whose position and energy are matched 
with extrapolated charged tracks and use tracker information for such particles;  
consider remaining clusters as neutral partiles and use calorimeter information.
The PandoraPFA\cite{pandoraPFA} uses the similar approach 
but introduced algorithm of re-clustering to disconnect merged clusters
or reconnect divided clusters, resulting better performance.
Another approach includes the algorithm to use the charged track information as a seed of 
the calorimeter clustering\cite{track-based-clustering}.

The performance of the GLD-PFA has been studied using the $Z$ pole events 
where $Z$ decays to $u$, $d$ or $s$ quarks only. 
The distribution of the observed particle energy tends to have two-gaussian 
distribution, broader one being caused by a loss of particles due to imperfect acceptance.
$\sigma_{90}$ is introduced as a measure of the PFA performance.
$\sigma_{90}$ is defined as the RMS of samples containing 90\% of all samples.
The resultant performance is shown in Figure~\ref{fig-gldpfa} as a 
function of the jet angle.  In the central region of $|\cos\theta_{jet}|<0.9$,
the obtained resolution is consistent with the goal.

However, for higher energy jets, the resolution of GLD-PFA gets worse and not satisfactory.
On the otherhand, the PandoraPFA has successfully updated its algorithm recently and the resolution 
of about $30\%\sqrt{E}$ has been achieved for a jet of energy up to 100 GeV.
The jet energy dependence of the energy resolution ($\Delta E / E$) of Pandora PFA 
is shown in Figure~\ref{fig-pfacompare}.  Further improvements of the performance are anticipated 
because studies using a perfect PFA indicates that improvements in the resolution 
for high energy jets would be achievable.

The number of detector optimization studies have been performed with the PandoraPFA.
For example, Figure~\ref{jetnergy-vs-tpcradius} shows
how the jet energy resolution depends on TPC radius ( which is almost the same 
as the inner radius of calorimeter ) and magnetic field.
This study is suggesting that the resolution improves with 
increasing the magnetic field strength but the larger radius of the calorimeter is more 
important than the stronger magnetic field.

The dual readout calorimeter system of the 4th concept does not have 
longitudinal segmentation, thus the jet energy is determined 
mainly by the calorimeter after the jet clustering using the cone 
algorithm. The tracker information is used to correct 
low $p_t$ tracks. The energy resolution of about $40\%\sqrt{E}$
has been reported\cite{ref-4thjet}.

\subsection{Integrated Physics Performance}

\begin{figure}
\begin{minipage}{0.51\textwidth}
\centerline{\includegraphics[width=\columnwidth]{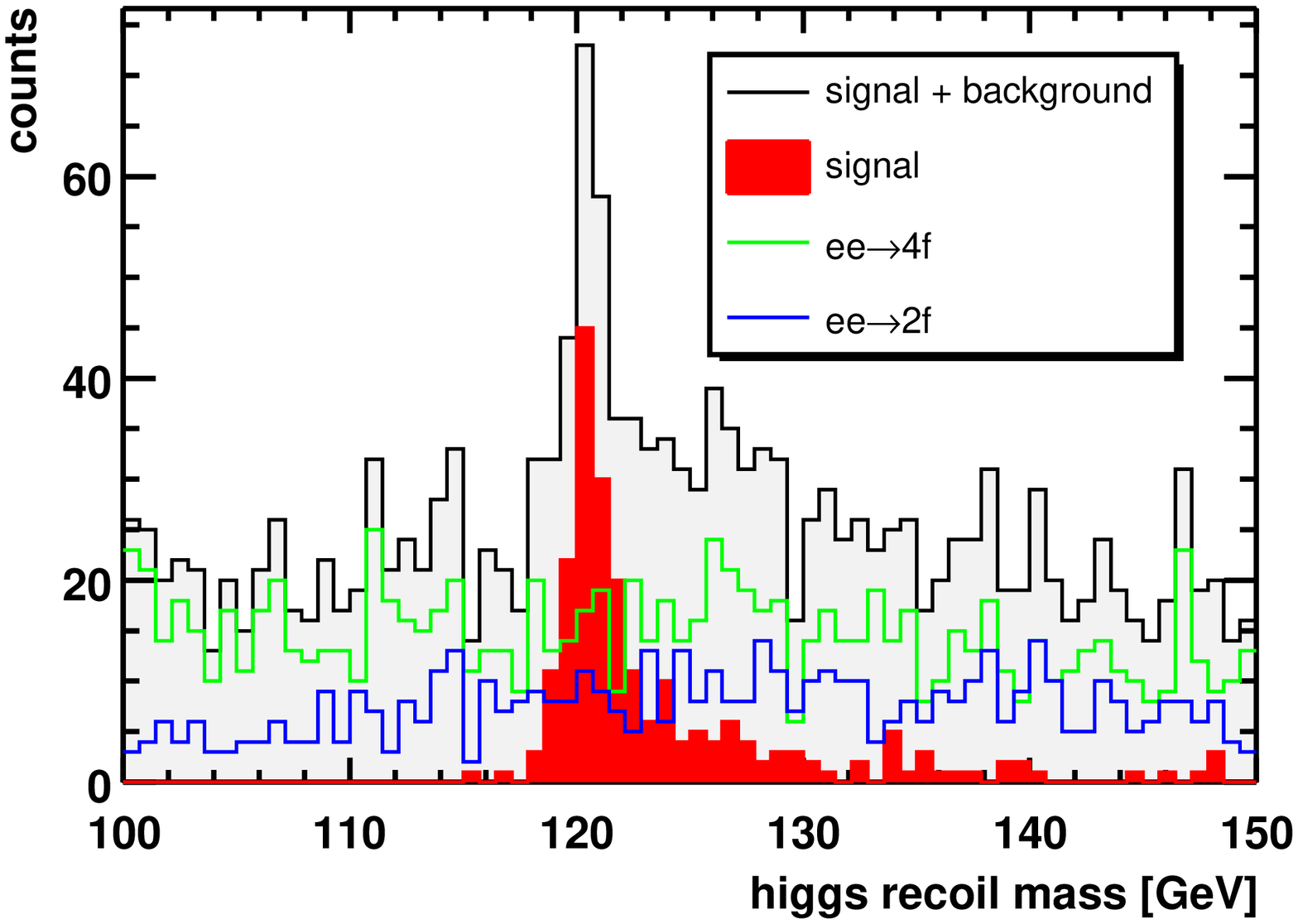}}
\begin{small}
\caption{{\it Recoil mass spectrum reconstructed for a 120~GeV Higgs, 
with full background simulation.}}
\label{fig:Higgs_recoil}
\end{small}
\end{minipage}
\hspace{0.01\textwidth}
\begin{minipage}{0.45\textwidth}
\centerline{\includegraphics[width=\columnwidth]{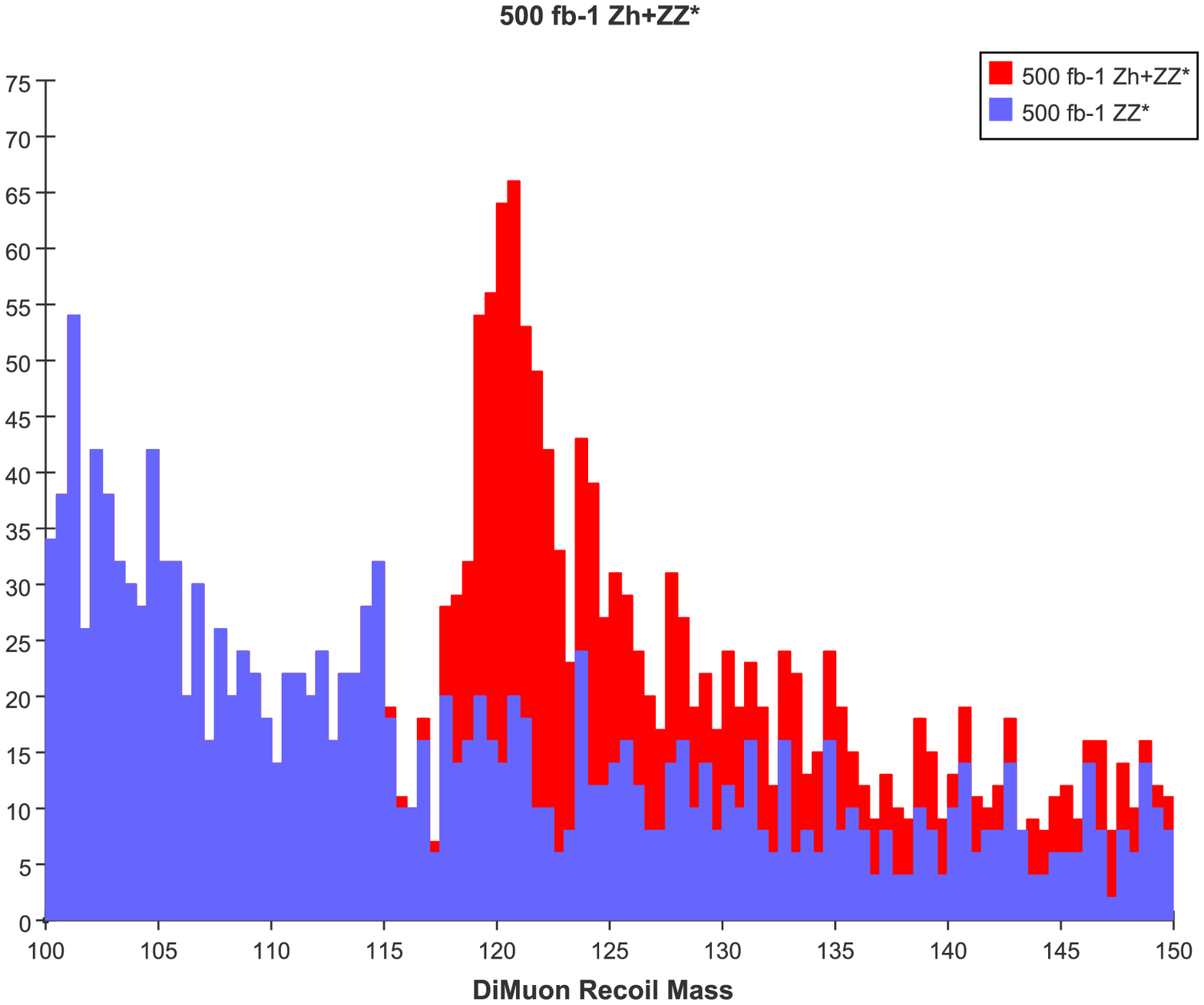}}
\begin{small}
\caption{{\it Di-muon recoil mass for $ZZ^*$ background (blue) and $ZH$ signal
    plus background (red) for centrally produced muons. 
}}
\label{fig:dimuonRecoilMass}
\end{small}
\end{minipage}
\end{figure}

In this paper, studies on the Higgs recoil mass measurement 
and on the $\nu\bar{\nu}b\bar{b}$ chanel of the Higgsstrahlung process are 
presented. A few more physics studies are described in the DCR.
The scope of the studies in this section is rather limited and 
does not cover the full physics potential of ILC. 
The purpose of this section is to
illustrate the level of maturity of both the understanding of the
detectors and of the reconstruction and analysis algorithms. 
Especially, development of the particle flow algorithms is still advancing rapidly. 
Therefore results presented in the following should be
interpreted as a snapshot of an ongoing development, where significant further improvements
can be expected in coming years.

The one of the most challenging reactions for the tracking system of the ILC detector 
is the measurement of the Higgs mass using the recoil mass technique.   
LDC has studied both $H\mu\mu$ and $Hee$ final states of $e^+e^-\rightarrow ZH$ process 
near threshold ( $\sqrt{s} \sim 250 $ GeV), including background processes 
of 4 fermions and 2 fermions final states. 
Based on a data sample equivalent of 50 fb$^{-1}$, a signal from the
Higgs has been reconstructed  as shown in Figure~\ref{fig:Higgs_recoil}.
From a simple fit to the mass distribution, 
the error of Higgs mass measurement is estimated to about 70 MeV 
and the relative cross section error being 8\%.

A similar analysis has been performed in the context of the SiD detector concept, 
at a center of mass energy of 350~GeV. 
The analysis fully simulated the machine background events as well. 
The background events have been combined with the signal events 
at the Monte Carlo hit level prior to digitization, 
then fed into a full track reconstruction code.  
Requiring two muons with momentum greater than 20~GeV, 
events whose invariant mass of the two-muon system is consisten with $Z$ were selected.
Figure~\ref{fig:dimuonRecoilMass} shows the recoil mass distribution
for the $ZZ^*$ background in blue and $ZH$ signal plus background in
red.
The precision of the Higgs mass from this measurement, based on a comparison between
the mass distribution reconstructed and template Monte Carlo distributions, is
estimated to be 135~MeV. Taking in to account the larger center of mass 
energy of this analysis, the result is consistent with the previous analysis.

\vspace{8pt}

\begin{wrapfigure}{r}{0.50\textwidth}
\centerline{\includegraphics[width=0.44\textwidth]{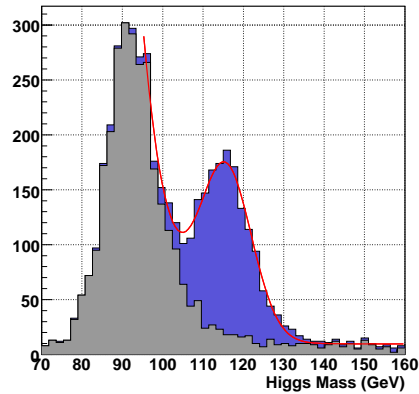}}
\begin{small}
\caption{{\it Reconstructed mass spectrum for Higgs candidates in the 
$ZH\rightarrow \nu \bar \nu b \bar b$ decay.}}
\label{fig:physics_ZHnnbb}
\end{small}
\end{wrapfigure}

GLD has studied the process,
$e^+e^- \rightarrow ZH$ at the center of mass energy of 350 GeV, where $Z$ decays to invisibly and $H$ 
decays to jets and the Higgs mass is 120 GeV/c$^2$.
In this case, compared to the four-jet mode, 
a beam energy constraint does not work for improve measurements
due to the missing particles. But there is no ambiguity in 
the mass measurement due to exchanges of colored particles in 
the final state because all of visible particles stem from the Higgs decay.
Thus high-performance PFA measurements is crucial 
for a good measurement.  $e^+e^- \rightarrow ZZ$ is the major 
background process and an excellent vertex detector is 
a key to reject them by discarding 
non-$b$ quark jets.

The preliminary result of GLD is shown in Figure~\ref{fig:physics_ZHnnbb}.
The analysis was based on a Monte Carlo sample of 200 fb$^{-1}$.
The effects of beamstrahlung as well as bremsstrahlung were included 
in the event generation. The Higgs signal is clearly seen 
above backgrounds, while further improvements of PFA performance is awaited
to achieve the signal width consistent with $30\%\sqrt{E}$.

\subsection{The case for two detectors}
Two complementary detectors are crucial for ILC, because 
it offers competing experiments, cross checking of results
and scientific redundancy for precision measurements at the level 
which is not created by more than one analysis teams for one detector;
significant increase of the scientific productivity, despite the spliting of the ILC luminosity;
maximal participation of the global particle physics community; 
the backup if one detector needs significant down time.
There are numerous historical examples where complementary 
experiments were critical. 
Further arguments will be find in Ref.\cite{twodetectors}.

\subsection{Costs}
The Costing Panel has been formed by WWS  to estimate costs of each 
concept by a common approach.  They have estimated the costs in light of the GDE costing rule and 
attempt to identify breakdown and cost drivers.  The cost breakdowns are different 
among concepts depending on how to categorize items, for example, 
a separation or inclusion of M\&S and man power costs. But, as naturally expected, 
calorimeters and magnets are the cost drivers.  Overall, there is a 
reasonable agreement among estimates by GLD, LDC and SiD and the total cost
lies in the range of $400\sim 500$ M\$ with about 20\% error.

\subsection{Options}

The one option is GigaZ, which aims to run at $Z$ pole energy with 
a luminosity of $\sim 4\times 10^{33}$cm$^{-2}$s$^{-1}$ and 
accumulate $10^9$ $Z$ events in one year.  Despite the high event rate,
the event overlap probability is less than 1\% and not a problem.
Challenges are to run with a polarized positron beam with a frequent change of its polarity 
in order to reduce systematics and measure the beam energy at precision less than $3\times 10^{-5}$.

The other is Photon Collider for experiments of $\gamma\gamma$ and $e\gamma$ collisions.
It provides a novel opportunity of physics such as studies of $\Gamma(H\rightarrow\gamma\gamma)$
and CP properties of the Higgs.  To make a $\gamma\gamma$ collision in the
ILC, the beam lines have to be modified to change the crossing angle from 14 mrad
to around 25 mrad.  
In addition, a $\gamma\gamma$ beam dump system has to be developed, to 
to deal with the $\gamma$ energy after collision:
the $\gamma$ beam is collimated and has the energy of about 50\% of 
the initial beam, but can not be steered or smeared out by 
magnets like $e^+/e^-$ beams.  
For Photon Collider experiments, 
near beam line components of detectors has to be modified to open a space to inject 
a laser light and to extract $\gamma$ beams. Additional space in 
a detector hall may be necessary for a laser optical cavity.

\section{Comments}
Editors appreciate for your patient reading of the draft and sending us valuable comments.  
We have received many technical comments, which will be included in the next version to be released 
after the workshop. There are another class of comments, where inputs from community 
are crucial.

One is regarding the goal of the jet energy measurement.
It has been set as $\Delta E/E \sim 30\%/\sqrt{E} \oplus const.$,
where the constant terms are usually neglected.
This goal is to achieve a jet-pair invariant mass resolution
 ($\Delta M_{12}/M_{12}$)
which is sufficient to separate $W$ and $Z$ in their hadronic decay modes.
The mass resolution of the jet pair is approximated,
in terms of the jet energy resolution($\Delta E_{i}/E_{i};i=1,2$), as 
\begin{equation}
\frac{\Delta M_{12}}{M_{12}}\sim \frac{1}{2}\left( 
\frac{\Delta E_1}{E_1} \oplus \frac{\Delta E_2}{E_2} \right),
\end{equation}
where the mass of the jets 
and the error of the angle between the jets are neglected.
For higher enrgy jets, the jet energy resolution is dominated by 
the constant term which would be mainly determined by a limitation of the PFA performance.
Therefore, it would be more appropriate to express the goal of the jet energy 
resolution in terms of $\Delta E/E$ rather than the coefficient 
of the stocastic term.
On the other hand, physics studies has been carried out assuming 
the formula, $\Delta E/E \sim \alpha/\sqrt{E}$ and studies 
assuming constant $\Delta E/E$ are yet to be done. The PFA performances 
will improve time to time and conservative opinions to keep 
the original arguments for the DCR have been made. 

One another issue is regarding the momentum resolution:
what do we gain by having the resolution which is significantly 
better than the original goal of $1\times 10^{-3} \oplus 5\times 10^{-5} p_t$ ?
If the di-lepton recoil mass of the process, $e^+e^- \rightarrow ZH$, is measured 
at $\sqrt{s}=350$ GeV for $M_h=120$ GeV, the resolution improves 
with better momentum resolution.  On the otherhand, as long as this measurement
is concerned, much better performance is obtained if measured just above 
the threshold. 

The statement in the draft DCR
will be rephrased taking account these arguments.
\section{Summary}
The overview of the draft detector DCR is presented.  The detector DCR describes
detector designs,
R\&Ds on detector technologies, and expected performances, aiming to make 
the case for the ILC detectors.

The author of the DCR consists of those who have participated in the detector concept studies, 
linear collider detector R\&D or have an interest in the physics and detectors for ILC.
Those who are qualified  are invited and encouraged to sign the DCR.  
The web page has been prepared for the sign up.

The draft is open to the public at \verb+http://www.linearcollider.org/wiki+ 
and comments from the community is welcomed.
The DCR Review Panel has been formed by WWS.  Preliminary comments from the panel 
is due by the end of LCWS2007 and the final report is expected by the beginning of July.
Taking into accounts these comments, the DCR is scheduled to submit to ILCSC in August.

\begin{center}
{\bf Acknowledgments}
\end{center}
The author wish to thank co-editors of the detector DCR whose help is indispensable 
for this talk: Ties Behnke, Chris Damerell, John Jaros. Moreover, 
the author would like to appreciate many colleagues of the ILC community whose work has been 
reported in the detector DCR.  This work is partially supported
by the Creative Scientific Research Grant
No. 18GS0202 of the Japan Society for Promotion of Science.


%

\begin{footnotesize}

\end{footnotesize}


\begin{thebibliography}{99}
\bibitem{DetectorRDR}
The Detector RDR is released as the detector volume of RDR this summer.
It is now available at the RDR web site\cite{RDR} 
together with other volumes 
on Executive Summary, Physics and Accelerator.
Details on ILC detectors will be found  
in the detector volume of RDR.
\end{thebibliography}
\end{document}